\documentclass{ws-mpla}

\begin{document}

\markboth{Y. Ping et al.} {Correspondence Between DGP Brane
Cosmology and $5D$ Ricci-flat Cosmology}

%%%%%%%%%%%%%%%%%%%%% Publisher's Area please ignore %%%%%%%%%%%%%%%
%
\catchline{}{}{}{}{}
%
%%%%%%%%%%%%%%%%%%%%%%%%%%%%%%%%%%%%%%%%%%%%%%%%%%%%%%%%%%%%%%%%%%%%
\title{Correspondence Between DGP Brane
Cosmology and $5D$ Ricci-flat Cosmology}

\author{YONGLI PING$^{\ast}$, LIXIN XU and HONGYA LIU$^{\dag}$}

\address{School of Physics and Optoelectronic Technology,\\ Dalian
University of Technology, Dalian, Liaoning 116024,
P.R.China\\$^{\ast}$ylping@student.dlut.edu.cn\\
$^{\dag}$hyliu@dlut.edu.cn}

\maketitle

\pub{Received }{Revised }

\begin{abstract}
We discuss the correspondence between the DGP brane cosmology and
$5D$ Ricci-flat cosmology by letting their metrics equal each other.
By this correspondence, a specific geometrical property of the
arbitrary integral constant $I$ in DGP metric is given and it is
related to the curvature of $5D$ bulk. At the same time, the
relation of arbitrary functions $\mu$ and $\nu$ in a class of
Ricci-flat solutions is obtained from DGP brane metric.
\end{abstract}

 \ccode{PACS numbers:
04.50.+h, 98.80.-k, 02.40.-k}

\section{Introduction}

An increasing number of people believe that our universe is a $5D$
spacetime which is a $4D$ spacetime with an extra dimension, such as
brane theory\cite{Arkani-Hamed,Arkani-Hamed2,Arkani-Hamed3,Horava,E.
Witten,Horava2,Randall1,Randall} and Space-Time-Matter (STM)
theory.\cite{Wesson99,O+W} In the brane world model, the gravity can
freely propagate in all dimensions, while the standard matter
particles and forces are confined on the 3-brane. The $5D$ brane
world with modified gravity is proposed by Dvali, Gabadadze and
Porrati (DGP).\cite{Dvali,Dvali2,Dvali3} There are the brane and
bulk Einstein terms in the action of DGP model. It was shown that
the DGP model allows for an embedding of the standard Friedmann
cosmology in the sense that the cosmological evolution of the
background metric on the brane can entirely be described by the
standard Friedmann equation plus energy conservation on the
brane.\cite{Dick,Dick2} And Dick gives an exact metric of $5D$ bulk.
Moreover DGP brane model is used to explain accelerating expansion
of universe.\cite{Deffayet,Deffayet2,Deffayet3,Alcaniz,D. Jain,A.
Lue} A comprehensive review on DGP cosmology is dished up in
Ref.~\refcite{Lue}.

In STM theory, the $5D$ manifold is Ricci-flat with $R_{AB}=0$ while
the $4D$ matter is induced from the $5D$ vacuum. This theory is
supported by the Campbell-Magaard theorem\cite{Campbell,Magaard}
which states that any analytic solution of Einstein's equations in
$N$-dimensions can be locally embedded in a $(N+1)$-dimensional
Ricci-flat manifold. A class of $5D$ Ricci-flat cosmological
solutions was firstly presented by Liu and Mashhoon\cite{L+M95} and
restudied latter by Liu and Wesson.\cite{L+W01} This class of
solutions is algebraically rich because it contains two arbitrary
functions of the time $t$. the solutions are utilized in
cosmology\cite{X-W,W-X,XU,CHANG,Liko,L-L,X-Z,ZHANG,PING} and are
relate to the brane model.\cite{Seahra,J. Ponce de Leon,Liu}

In this paper, we discuss the correspondence between DGP brane
cosmology and $5D$ Ricci-flat cosmology. By studying the solutions
in DGP model and $5D$ Ricci-flat cosmology, a clear geometrical
property of the Integral constant $I$ is obtained in DGP brane, and
also a constraint is given on a class of solutions in STM theory.
Then the evolution of the scale factor $a(y,t)$ and the condition
where $a(y,t)$ has a bounce are discussed.

\section{DGP Brane cosmology and $5D$ Ricci-Flat cosmology}

We consider a general five-dimensional spacetime metric. Because we
are interested in cosmological solutions, the metric is taken as
\begin{equation}
dS^2=-n^2(y,t)dt^2+a^2(y,t)\gamma_{ij}dx^{i}dx^{j}+b^2(y,t)dy^2,
\end{equation}
where, $\gamma_{ij}$ is a maximally symmetric $3D$ metric where
$k=0,\pm1$ parameterizes the spatial curvature. Adopting the
Gaussian normal system  gauge $b(y,t)=1$, the $5D$ Einstein tensors
$\tilde{G}_{AB}$ are
\begin{eqnarray}
\tilde{G}_{00}&=&3n^2\left(\frac{\dot{a}^2}{n^2a^2}-\frac{a'^2}{a^2}+\frac{k}{a^2}\right)-3n^2\frac{a''}{a}\label{G00},\\
\tilde{G}_{ij}&=&\left(\frac{a'^2}{a^2}-\frac{\dot{a}^2}{n^2a^2}-\frac{k}{a^2}\right)\gamma_{ij}
+2\left(\frac{a''}{a}
+\frac{n'a'}{na}-\frac{\ddot{a}}{n^2a}+\frac{\dot{n}\dot{a}}{n^3a}\right)\gamma_{ij}+\frac{n''}{n}\gamma_{ij}\label{Gij},\\
\tilde{G}_{05}&=&3\left(\frac{n'\dot{a}}{na}-\frac{\dot{a}'}{a}\right)\label{G05},\\
\tilde{G}_{55}&=&3\left(\frac{a'^2}{a^2}-\frac{\dot{a}^2}{n^2a^2}-\frac{k}{a^2}\right)
+3\left(\frac{n'a'}{na}+\frac{\dot{n}\dot{a}}{n^3a}-\frac{\ddot{a}}{n^2a}\right)\label{G55}.
\end{eqnarray}

Firstly, in DGP brane cosmology, the five-dimensional Einstein
equations take the form as
\begin{equation}
\tilde{G}_{AB}=\tilde{R}_{AB}-\frac{1}{2}\tilde{R}\tilde{g}_{AB}=\kappa_{(5)}^2\tilde{S}_{AB},\label{GAB}
\end{equation}
where $\tilde{R}_{AB}$ is the $5D$ Ricci tensor,
$\tilde{R}=\tilde{g}^{AB}\tilde{R}_{AB}$ is the scalar curvature,
$\kappa_{(5)}$ is related to the $5D$ Newton's constant $G_{(5)}$
and $\kappa_{(5)}^2=8\pi{G_{(5)}}=M_{(5)}^{-3}$ where $M_{(5)}$ is
the $5D$ Planck mass. The stress-energy-momentum tensor $\tilde{S}$
contains three parts,
\begin{equation}
\tilde{S}^A_B=\check{T}^A_B|_{bluk}+T^A_B|_{brane}+\tilde{U}_{AB},
\end{equation}
where $\check{T}^A_B|_{bluk}$ and $T^A_B|_{brane}$ are the energy
momentum tensor of bulk and brane respectively and $\tilde{U}_{AB}$
is contribution coming from the scalar curvature of the brane.
$\check{T}^A_B|_{bluk}$ and $T^A_B|_{brane}$ are written as
\begin{eqnarray}
\check{T}^A_B|_{bluk}=diag(-\rho_B,P_B,P_B,P_B,P_B),\\
T^A_B|_{brane}={\delta(y)}diag(-\rho_b,p_b,p_b,p_b,0),
\end{eqnarray}
where the energy density $\rho_B$ and pressure $P_B$ of the bulk are
independent of the coordinate $y$; the energy density $\rho_b$ and
pressure $p_b$ are the function of time.

For $\tilde{T}_{05}=0$, Eq. (\ref{G05}) implies
\begin{equation}
\frac{n'}{n}=\frac{\dot{a}'}{\dot{a}}.\label{na}
\end{equation}
From Eqs. (\ref{G00}), (\ref{G55}) and (\ref{GAB}), they are
obtained\cite{Binetruy}
\begin{eqnarray}
\frac{\partial}{\partial{y}}\left(\frac{\dot{a}^2}{n^2}a^2-a'^2a^2+ka^2\right)&=&\frac{2}{3n^2}a'a^3\kappa^2{\rho_B},\label{G-rho}\\
\frac{\partial}{\partial{t}}\left(\frac{\dot{a}^2}{n^2}a^2-a'^2a^2+ka^2\right)&=&\frac{2}{3}\dot{a}a^3\kappa^2{P_B}.\label{G-P}
\end{eqnarray}
Assuming there is nothing in the bulk, they have $\rho_B=0$ and
$P_B=0$. Then, Eqs. (\ref{G-rho}) and (\ref{G-P}) are rewritten
as\cite{Dick,Dick2}
\begin{eqnarray}
\frac{\partial}{\partial{y}}\left(\frac{\dot{a}^2}{n^2}a^2-a'^2a^2+ka^2\right)&=&0,\\
\frac{\partial}{\partial{t}}\left(\frac{\dot{a}^2}{n^2}a^2-a'^2a^2+ka^2\right)&=&0,
\end{eqnarray}
and
\begin{eqnarray}
I^+&=&\left(\frac{\dot{a}^2}{n^2}-a'^2+k\right)a^2 \bigg |_{y>0}\label{I+},\\
I^-&=&\left(\frac{\dot{a}^2}{n^2}-a'^2+k\right)a^2 \bigg |_{y<0},
\end{eqnarray}
are two constants. In Ref.~[\refcite{Dick,Dick2}], the author find
that the standard Friedmann equation holds on the brane is
equivalent to the smoothness condition,
\begin{equation}
\lim_{\epsilon\rightarrow{+0}}a'
\big|_{y=\epsilon}=\lim_{\epsilon\rightarrow{+0}}a'
\big|_{y=-\epsilon},
\end{equation}
which means that $a$ is smooth across the brane. So, this leads to
$I^+=I^-=I$.

In order to simplify the previous equations, the gauge $n(0,t)=1$ is
adopted. Then, from Eq. (\ref{na}), $n(y,t)$ is obtained as
\begin{equation}
n(y,t)=\frac{\dot{a}(y,t)}{\dot{a}(0,t)}.\label{n}
\end{equation}
By substituting $n(y,t)$ into Eq. (\ref{I+}), the constant $I$ is
rewritten as
\begin{equation}
I=\left(\dot{a}^2(0,t)-a'^2(y,t)+k\right)a^2(y,t).\label{I}
\end{equation}
Choosing $a'(y,t)>0$ for the sign of $y$ in the direction of
increasing scale factor, from Eqs. (\ref{I}) and (\ref{n}), they are
obtained as
\begin{eqnarray}
a^2(y,t)&=&a^2(0,t)+(\dot{a}^2(0,t)+k)y^2+2\sqrt{(\dot{a}^2(0,t)+k)a^2(0,t)-I}y,\label{a(y,t)}\\
n(y,t)&=&\left[a(0,t)+\ddot{a}(0,t)y^2+a(0,t)y\frac{a(0,t)\ddot{a}(0,t)+\dot{a}^2(0,t)+k}{\sqrt{(\dot{a}^2(0,t)+k)a^2(0,t)-I}}\right]\nonumber\\
&&\times\left[a^2(0,t)+(\dot{a}^2(0,t)+k)y^2+2\sqrt{(\dot{a}^2(0,t)+k)a^2(0,t)-I}y\right]^{-1/2}.\label{n(y,t)}
\end{eqnarray}
The metric components lead to an exact solution of $5D$ spacetime.
We can find the solution is similar to the bounce solution which is
derived from STM theory.\cite{L+M95,L+W01}

In the STM theory for the $5D$ Ricci-flat cosmology, the Ricci
tensor are $\tilde{R}_{AB}=0$. Therefore, in term of the $5D$
Einstein tensor
$\tilde{G}_{AB}=\tilde{R}_{AB}-\frac{1}{2}\tilde{R}\tilde{g}_{AB}$,
they are
\begin{equation}
\tilde{G}_{AB}=0.\label{G0}
\end{equation}
The $4D$ Einstein field equations are given as
\begin{equation}
G_{\mu\nu}=T_{\mu\nu}.\label{G4}
\end{equation}
The central idea of STM theory is that (\ref{G4}) is a subset of
(\ref{G0}) with the induced $4D$ energy-momentum tensor $T_{\mu\nu}$
which has the classical properties of matter.

A class of $5D$ Ricci-flat cosmological solutions in STM theory
reads\cite{L+W01}
\begin{equation}
dS^{2}=B^{2}dt^{2}-A^{2}\left( \frac{dr^{2}}{1-kr^{2}}+r^{2}d\Omega
^{2}\right) -dy^{2}, \label{line element}
\end{equation}%
\begin{equation}
A^{2}=\left( \mu ^{2}+k\right) y^{2}+2{\nu }y+\frac{\nu ^{2}+K}{\mu
^{2}+k} ,  \label{A}
\end{equation}%
\begin{equation}
B=\frac{1}{\mu }\frac{\partial A}{\partial t}\equiv
\frac{\dot{A}}{\mu },  \label{B}
\end{equation}%
where $d\Omega ^{2}=d\theta ^{2}+\sin^2 \theta d\psi ^{2}$, $\mu
=\mu \left( t\right) $ and $\nu =\nu \left( t\right) $ are two
arbitrary functions of time $t$, $k$ is the 3D curvature index
($k=\pm 1,0$), and $K$ is a constant. Because the $5D$ manifold
(\ref{line element})-(\ref{B}) is Ricci-flat, we have $I_{1}\equiv
R=0$, $I_{2}\equiv R^{AB}R_{AB}=0$, and%
\begin{equation}
I_{3}\equiv R^{ABCD}R_{ABCD}=\frac{72K^{2}}{A^{8}} ,  \label{I-3}
\end{equation}%
so $K$ is related to the $5D$ curvature.d

Comparing (\ref{a(y,t)}) with (\ref{A}), we find if
\begin{eqnarray}
\mu&=&\dot{a}(0,t),\label{mu-a}\\
\nu&=&\sqrt{(\dot{a}^2(0,t)+k)a^2(0,t)-I},\label{nu-a}\\
a^2(0,t)&=&\frac{\nu^2+K}{\mu^2+k}\label{a-mu},
\end{eqnarray}
the bulk solutions will be the same as the Ricci-flat solutions.
Meanwhile, we can obtain $I=K$. In brane theory, the $I$ in
(\ref{a(y,t)}) is only an arbitrary integral constant, while $K$ is
related to the $5D$ curvature in the Ricci-Flat cosmology.
Therefore, the constant $I$ has a new geometrical property in the
brane theory. The constant $I$ is related to the $5D$ curvature of
the $5D$ bulk. In the Ricci-flat solutions, by Eqs. (\ref{mu-a}) and
(\ref{a-mu}), the relation of arbitrary functions $\mu$ and $\nu$ is
given
\begin{equation}
\mu(\nu^2+K)^{1/2}(\mu^2+k)^{3/2}
=\nu\dot{\nu}(\mu^2+k)-\mu\dot{\mu}(\nu^2+K).\label{mu-nu}
\end{equation}
For $K=0$, the $5D$ Kretschmann invariant in (\ref{I-3}) will be
$I_1=I_2=I_3=0$. In the braneworld, $K=0$ leads to $I=0$. In this
case the components of metric in Eqs. (\ref{a(y,t)}) and
(\ref{n(y,t)}) are
\begin{eqnarray}
a(y,t)&=&a(0,t)+\sqrt{\dot{a}^2(0,t)+k}y,\label{a0}\\
n(y,t)&=&1+\frac{\ddot{a}(0,t)}{\sqrt{\dot{a}^2(0,t)+k}}y.\label{n0}
\end{eqnarray}
When $k=0$, the Eqs. (\ref{a0}) and (\ref{n0}) become
\begin{eqnarray}
a(y,t)&=&a(0,t)+\dot{a}(0,t)y,\label{a00}\\
n(y,t)&=&1+\frac{\ddot{a}(0,t)}{\dot{a}(0,t)}y.\label{n00}
\end{eqnarray}
With the relations $\mu=\dot{a}(0,t)$ and $a(0,t)=\nu/\mu$ where
$k=K=0$, Eqs. (\ref{a00}) and (\ref{n00}) are expressed with $\mu$
and $\nu$ as
\begin{eqnarray}
a(y,t)&=&\frac{\nu}{\mu}+\mu{y},\\
n(y,t)&=&1+\frac{\dot{\mu}}{\mu}y.
\end{eqnarray}
When $k=0$ and $K=0$, the relation Eq. (\ref{mu-nu}) becomes
\begin{equation}
\mu^3=\dot{\nu}\mu-\dot{\mu}\nu,
\end{equation}
and the components of the $5D$ Ricci-flat metric (\ref{A}) and
(\ref{B}) are rewritten as
\begin{eqnarray}
A(y,t)&=&\frac{\nu}{\mu}+\mu{y},\\
B(y,t)&=&1+\frac{\dot{\mu}}{\mu}y.
\end{eqnarray}
Obviously, they have the same forms as the ones in brane. However
they have different motivation. The main difference is that authors
adopt different points of view regarding the matter content and
dynamics of $4D$ spacetime in brane and STM theory.

\section{The evolution of scale factor $a(y,t)$}

\begin{figure}
\begin{center}
\includegraphics[angle=0,width=3.0in]{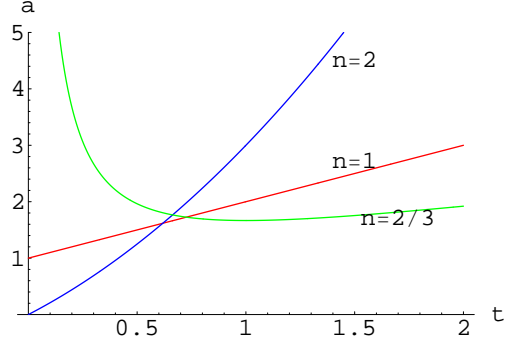}
\end{center}
\caption{Evolution of the scale factor $a(y,t)$ with $y=1$,
$n=2/3,1,2$.}\label{a-plot}
\end{figure}

In the DGP model, only one brane is contained and put at the
position of $y=0$, where the evolution of brane universe is
described by the scale factor $a(0,t)$. And the bulk solution is
derived as the equation (\ref{a(y,t)}). However, in the STM theory,
our universe is a hypersurface $y=constant$. So, there is a
correspondence between the solution of DGP model and STM theory. In
details, once the evolution $a(0,t)$ on the DGP brane is set, the
evolution of STM universe is described by the scale factor $a(y,t)$.
For example, we let $a(0,t)=t^n$ on the brane, and then derive the
scale factor in the STM theory as
\begin{eqnarray}
a(y,t)&=&t^n+nt^{n-1}y,\label{atn}\\
n(y,t)&=&1+(n-1)t^{-1}y.\label{ntn}
\end{eqnarray}
We calculate the first and second derivative of $a(y,t)$ with
respect to time $t$ and get
\begin{eqnarray}
\dot{a}(y,t)&=&nt^{n-1}+n(n-1)t^{n-2}y,\label{a.}\\
\ddot{a}(y,t)&=&n(n-1)t^{n-2}+n(n-1)(n-2)t^{n-3}y.\label{a..}
\end{eqnarray}
Considering the Eqs. (\ref{a.}) and (\ref{a..}), if there is a
bounce on $y=constant$ supersurface in STM cosmology, the necessary
condition that $a(y,t)$ has the extremum is $\dot{a}=0$ and
$\ddot{a}(0,t)>0$. In contrast, a non-bounce cosmology is
$\dot{a}\neq0$ and $\ddot{a}\geq0$. When $\dot{a}=0$, from Eq.
(\ref{a.}), we have
\begin{equation}
t_b=-(n-1)y.
\end{equation}
For $\ddot{a}>0$, from Eq. (\ref{a..}), we let $y>0$ and get
$0<n<1$. Therefore, $a(y,t)$ should have minimum when $0<n<1$ and we
name it as a bounce. On the contrary when $t\neq-(n-1)y$, there is
non-bounce. We plot the evolution of the scale factor with $y=1$
when $n=1$, $n>1$ and $n<1$ respectively. Fig.\ref{a-plot} shows the
evolution of the scale factor $a(y,t)$ with different values of $n$
and $y=1$.

\section{Conclusions}

In this paper, we have studied the correspondence between DGP brane
cosmology and $5D$ Ricci-flat cosmology by researching their exact
solutions. The solutions given by Dick in DGP brane are very similar
to the bounce solutions in Ricci-flat cosmology. Contrasting these
two solutions we find these two solutions have the same form when
the constant $I$ in DGP brane and $K$ in Ricci-flat cosmology
satisfy $I=K$ and the arbitrary functions $\mu$ and $\nu$ satisfy
(\ref{mu-nu}). Therefore, in this way, the arbitrary integral
constant $I$ is endowed with specific geometrical property and it is
related to the curvature of $5D$ bulk. At the same time, the
relation of the arbitrary function $\mu$ and $\nu$ in Ricci-flat
cosmology is obtained. Finally, the evolution of the scale factor
$a(y,t)$ in STM cosmology is discussed by giving $a(0,t)$ on the DGP
brane. Let $a(0,t)=t^n$ in the brane, the evolution of $a(y,t)$ in
STM cosmology is determined by $n$ with $y=constant$. We obtain the
necessary condition which $a(y,t)$ has the bounce is $0<n<1$ and
$y>0$. In Fig.\ref{a-plot}, the evolution of the scale factor
$a(y,t)$ with $y=1$ is shown with different values of $n$, where
$n=1$ corresponds to an uniform speed expansion universe, $n\geq1$
checks with no bounce, while $n<1$ squares with a bounce universe.

\section*{Acknowledgments}
This work was supported by NSF (10573003), NSF (10647110), NBRP
(2003CB716300) of P. R. China and DUT 893321.

\end{document}